\documentclass[11pt,a4paper,titlepage]{article}
\usepackage{cite}
\usepackage{color}
\usepackage{mathtools}
\usepackage{amsthm,amstext,amscd}
\usepackage{epsfig,graphicx,bm}
\usepackage{epstopdf, epsf}
\usepackage{dcolumn}
\usepackage{tensor}
\usepackage{datetime}
\usepackage{textalpha}
\newdateformat{monthyeardate}{%
  \monthname[\THEMONTH], \THEYEAR}
  
\usepackage{graphicx,caption,subcaption}
\usepackage[colorlinks,
           linkcolor = blue,		
           citecolor = blue,
           ]{hyperref}

\newcommand{\be}{\begin{equation}}
\newcommand{\ee}{\end{equation}}

\newcommand{\bse}{\begin{subequations}}
\newcommand{\ese}{\end{subequations}}
\newcommand{\bea}{\begin{eqnarray}}
\newcommand{\eea}{\end{eqnarray}}
\newcommand{\ba}{\begin{array}}
\newcommand{\ea}{\end{array}}
\newcommand{\bc}{\begin{center}}
\newcommand{\ec}{\end{center}}

\def\cl{{\cal L}}

\def\ba{{\bf A}}

\def\bc{{\bf C}}

\def\bo{{\bf O}}

\let\du=\du                     

\def\b{\beta}

\def\d{\delta}

\def\f{\phi}

\def\k{\kappa}

\def\p{\pi}

\def\z{\zeta}

\def\ve{\varepsilon}

\def\bo{{\raise-.3ex\hbox{\large$\Box$}}}               
\def\TH{{\raise.2ex\hbox{$\displaystyle \bigodot$}\mskip-4.7mu \llap H \;}}
\def\face{{\raise.2ex\hbox{$\displaystyle \bigodot$}\mskip-2.2mu \llap {$\ddot
        \smile$}}}                                      

\def\VEV#1{\left\langle #1\right\rangle}        
\def\abs#1{\left| #1\right|}                    
\def\leftrightarrowfill{$\mathsurround=0pt \mathord\leftarrow \mkern-6mu
        \cleaders\hbox{$\mkern-2mu \mathord- \mkern-2mu$}\hfill
        \mkern-6mu \mathord\rightarrow$}
\def\dvec#1{\vbox{\ialign{##\crcr
        \leftrightarrowfill\crcr\noalign{\kern-1pt\nointerlineskip}
        $\hfil\displaystyle{#1}\hfil$\crcr}}}           

\def\frac#1#2{{\textstyle{#1\over\vphantom2\smash{\raise.20ex
        \hbox{$\scriptstyle{#2}$}}}}}                   
\def\sfrac#1#2{{\vphantom1\smash{\lower.5ex\hbox{\small$#1$}}\over
        \vphantom1\smash{\raise.4ex\hbox{\small$#2$}}}} 
\def\bfrac#1#2{{\vphantom1\smash{\lower.5ex\hbox{$#1$}}\over
        \vphantom1\smash{\raise.3ex\hbox{$#2$}}}}       
\def\afrac#1#2{{\vphantom1\smash{\lower.5ex\hbox{$#1$}}\over#2}}    

\def\[{\lfloor{\hskip 0.35pt}\!\!\!\lceil}
\def\]{\rfloor{\hskip 0.35pt}\!\!\!\rceil}

\def\du#1#2{_{#1}{}^{#2}}

\def\fracmm#1#2{{{#1}\over{#2}}}

\def\low#1{{\raise -3pt\hbox{${\hskip 0.75pt}\!_{#1}$}}}

\newskip\humongous \humongous=0pt plus 1000pt minus 1000pt

\newif\ifdtup

\newcommand{\nbe}{\begin{equation*}}
\newcommand{\nee}{\end{equation*}}

\newcommand{\lb}{\label}

\begin{document}

\thispagestyle{empty}

{\hbox to\hsize{
\vbox{\noindent December 2025 \hfill IPMU19-0159 }}}

\noindent
\vskip2.0cm
\begin{center}

{\large\bf Single-field D-type inflation in the minimal supergravity  \\in light of Planck-ACT-SPT data}
\vglue.3in

Yermek Aldabergenov~${}^{a}$ and Sergei V. Ketov~${}^{b,c,d}$
\vglue.2in
${}^a$~Department of Physics, Fudan University, 220 Handan Road, Shanghai 200433, China\\
${}^b$~Faculty of Science, Tokyo Metropolitan University, 1-1 Minami-Osawa, Hachioji, Tokyo 192-0397, Japan \\
${}^c$~Interdisciplinary Laboratory,  Tomsk State University, 36 Lenin Avenue, Tomsk 634050, Russia\\
${}^d$~Kavli Institute for the Physics and Mathematics of the Universe (WPI),
\\The University of Tokyo Institutes for Advanced Study, 5-1-5 Kashiwanoha, Kashiwa, Chiba 277-8583, Japan \\
\vglue.2in
ayermek@fudan.edu.cn, ketov@tmu.ac.jp
\end{center}
\vglue.3in

\begin{center}
{\Large\bf Abstract}
\end{center}
\vglue.1in
\noindent  The minimal supergravity framework is applied to a construction of new D-type single-field models of inflation
in agreement with precision measurements of the cosmic microwave background radiation by Planck Collaboration,
BICEP/Keck Collaboration, Atacama Cosmology Telescope and South Pole Telescope. The inflaton potential, the power spectrum of scalar perturbations, the cosmological observables and the reconstruction procedure can be very simple when using the e-folds as the running variable. 

\newpage

\section{Introduction}

Single-field models can provide minimal and viable description of cosmological inflation in terms of a real canonical scalar field $\phi$ (called inflaton) having a potential $V(\phi)$ and minimally coupled to Einstein gravity, in good agreement with observations of the cosmic microwave background (CMB) radiation \cite{Martin:2024qnn}.  On the one hand, the single-field models are severely constrained by the CMB observations but, on the other hand, there are still many viable choices for the inflaton potential $V(\phi)$. This implies the need to complement the data-driven selection of $V(\phi)$ by fundamental theoretical principles and/or simplicity arguments.

One such fundamental principle suitable for high-scale inflation is given by local supersymmetry (we mean N=1 supersymmetry in four space-time dimensions) that implies general covariance and leads to supergravity. Inflation in supergravity is usually realized by embedding real inflaton into a chiral supermultipet, which leads to two physical scalars and related complications in describing inflation
by the F-term-generated scalar potential, see e.g., Ref.~\cite{Gong:2025pbi}. However, there is the alternative to the common F-type inflation models via embedding real inflaton into a massive vector supermultiplet and using the D-term-generated scalar potential in the minimal
supergravity having only one (real) physical scalar \cite{Farakos:2013cqa,Ferrara:2013wka,Aldabergenov:2017hvp}. 

Various CMB measurements with increasing precision done by Planck (P) \cite{Planck:2018jri}, 
BICEP/Keck \cite{BICEP:2021xfz}, Atacama Cosmology Telescope (ACT) \cite{ACT:2025tim}
and South Pole Telescope (SPT) \cite{SPT-3G:2025bzu} led to an explosion of new theoretical
models of inflation  \cite{Dioguardi:2025vci, Kallosh:2025rni, Gialamas:2025ofz, Aoki:2025wld, Berera:2025vsu,Brahma:2025dio, Gialamas:2025kef, Salvio:2025izr, Antoniadis:2025pfa, Dioguardi:2025mpp,Kuralkar:2025zxr,Gao:2025onc,He:2025bli,Drees:2025ngb,Haque:2025uri,Liu:2025qca,Addazi:2025qra, Byrnes:2025kit, Yogesh:2025wak, Kallosh:2025ijd, Kohri:2025lau, Ahmad:2025mul, Okada:2025lpl, Choudhury:2025vso, Han:2025cwk, Leontaris:2025hly, McDonald:2025tfp, Gao:2025viy, Wolf:2025ecy,Singh:2025uyr,Fu:2025ciy,Ahmed:2025eip} but even in the case of single-field slow-roll 
 inflation, the inflaton potential cannot be fixed.

We use the following observational constraints on the tilt $n_s$ of the power spectrum of scalar perturbations and the tensor-to-scalar ratio $r$:
\begin{gather}
    {\rm Planck:~~}n_s=0.9668\pm 0.0037~,~~~r<0.036~,\label{ns_Planck}\\
    {\rm P{\text +}ACT:~~}n_s=0.9752\pm 0.003~,~~~r<0.038~,\label{ns_P_ACT}
\end{gather}
where the Planck constraint on $n_s$ includes the lensing+BAO+BICEP/Keck 2015 data \cite{Planck:2018jri}, the Planck constraint on $r$ is based on the improved BICEP/Keck 2018 data \cite{BICEP:2021xfz}, the Planck+ACT constraint on $n_s$ is based on  the lensing+BAO data, and the Planck+ACT constraint on $r$ is based on the BICEP/Keck 2018 data. For all the constraints the pivot scale is $k_*=0.05~{\rm Mpc}^{-1}$. The SPT data implies  \cite{SPT-3G:2025bzu}
\begin{equation}\label{ns_SPA}
    {\rm SPA:~~}n_s=0.9684\pm 0.003~,
\end{equation}
where SPA=SPT+Planck+ACT. There is no improvement from the SPT data for the value of $r$ beyond the results of Planck or Planck+ACT.

As regards the running index $\alpha_s$ of the scalar tilt $n_s$, we use the observational
constraints \cite{ACT:2025tim},
\be \lb{alphas}
\alpha_s = 0.0062 \pm 0.0104~,
\ee 
i.e. the $\alpha_s$ is between $-0.004$ and $+0.017$ at 95\% C.L.

In this paper, we meet the CMB data given above in our single-field inflation models of the minimal supergravity.

The paper is organized as follows. In Sec.~2, we recall the basic notions of the power spectrum of
cosmological perturbations, the cosmological tilts, and their relation to CMB observations. Also, in Sec.~2 we illustrate our approach on the simplest examples without supersymmetry and supergravity
by emphasizing the dependence of the power spectrum and the cosmological tilts upon the e-folds of
inflation, in the slow-roll approximation. The D-type realisation of single-field inflation in the minimal supergravity framework is considered in Sec.~3 where new inflation models compatible with all CMB observations are also proposed and studied. Our conclusion is Sec.~4.

\section{Primordial power spectrum and cosmological tilts}

Primordial scalar (density) perturbations $\z$ and primordial  tensor perturbations (primordial gravitational waves) $g$ are described by a perturbed Friedmann-Lemaitre-Robertson-Walker (FLRW) metric,
\be \lb{perm}
ds^2 = dt^2 - a^2(t)\left( \d_{ij}+ h_{ij}(\vec{r})\right) dx^idx^j~~,\qquad i,j=1,2,3~,
\ee
where
\be \lb{pertur}
h_{ij}(\vec{r}) =2\z(\vec{r})\d_{ij} + \sum_{a=1,2} g^{(a)}(\vec{r})e^{(a)}_{ij}(\vec{r})~~,
\ee
and the basis tensors $e^{(a)}$ obey the relations $e^{i(a)}_{i}=0$, 
 $g^{(a)}_{,j}e^{j(a)}_{i}=0$ and $e^{(a)}_{ij}e^{ij(a)}=1$.

The primordial spectrum $P_{\z}(k)$ of scalar perturbations  is defined by the 2-point correlation function of scalar perturbations via Fourier transform,
\be \lb{hz}
\VEV{\z^2(\vec{r})}=\int dk \fracmm{P_{\z}(k)}{k}e^{ikr}~,
\ee
while the observed (Harrison-Zeldovich) spectrum of the CMB radiation is given by
\be P_{\z}(k)=2.21^{+0.07}_{-0.08} \times 10^{-9}
\left(\fracmm{k}{k_*}\right)^{n_s-1}~~,
\ee
with the pivot scale  $k_*=0.05~{\rm Mpc}^{-1}$ and the scalar tilt $n_s$ close (but not equal) to one.
The running spectrum $P_{\z}(k)$ is related to the (canonical) inflaton scalar potential $V_k(\phi)$ at the time $t_k$ as 
\be \lb{pvrel}
P_{\z}(k) =\fracmm{V_k^3}{12\pi^2{M^6_{\rm Pl}V'_k}^2}~~,
\ee
where the prime denotes the derivative with respect to $\phi$ and the subscript $k$ refers to the value of the scalar potential at $k=\dot a(t_k)$. The scalar tilt  (or the spectral slope) $n_s$ is related to the scalar potential $V_k$ as 
\be \lb{sslope}
n_s(k)-1 = \fracmm{d\ln P_{\z}}{d\ln k} =M^2_{\rm Pl}
\left( 2 \fracmm{{V_k}''}{V_k} - 3\fracmm{{{V_k}'}^2}{{V_k}^2}\right)~.
\ee

The power spectrum $P_g(k)$ of tensor perturbations is defined similarly to that in Eq.~(\ref{hz}) with the tensor tilt
\be \lb{tent}
n_g(k) = \fracmm{d\ln P_{g}}{d\ln k} =- M^2_{\rm Pl}
\left( \fracmm{{{V_k}'}^2}{{V_k}^2}\right)
\ee
and the tensor-to-scalar ratio
\be\lb{tsr}
r = \fracmm{P_{g}(k)}{P_{\z}(k)} =8\abs{n_g(k)}~~.
\ee

Instead of time $t$, scale $k$ or field $\phi$,  it is often more convenient  to choose the e-folds number $N$ as the {\it running} variable defined by  $N(k)=\ln\fracmm{k_f}{k}$ or
\be\lb{efoldsk}
dN = - \fracmm{dk}{k}~.
\ee
Then the inflaton equation of motion in the slow-roll (SR)  approximation takes the form
\be \lb{chva}
\left(\fracmm{d\f}{dN}\right)^2=M^2_{\rm Pl} \fracmm{d \ln V}{dN}
\ee 
and leads to the simple relation between the running tensor-to-scalar ratio $r(N)$ and the function $V(N)$ as
\be \lb{r}
 r(N) = 8 \fracmm{d\ln V}{dN}~.
\ee
Should the scalar potential $V$ have a plateau, $V=V_0 +\d V$, with $|\d V|\ll |V|$ and
$V_0=const.>0$,  Eq.~(\ref{r}) can be further simplified to
\be \lb{r2}
 r(N) = \fracmm{8}{V_0} \fracmm{d \d V}{dN}~.
\ee

A reconstruction of the scalar potential from the given power spectrum also takes the simple form with the e-folds $N$ as the running variable \cite{Hodges:1990bf},
\be \lb{hb1}
\fracmm{1}{V(N)} = - \fracmm{1}{12\pi^2 M^4_{\rm Pl}}\int \fracmm{dN}{P_{\z}(N)}~,
\ee
though one should keep in mind that the use of this equation implies knowing $P_{\z}(N)$ and the integration region outside the
SR approximation.

Let us take a simple ansatz  for the inflaton potental $V(N)$ in the form
\be \lb{try}
 V(N) = V_0 \fracmm{N}{N+N_0}
\ee
with the positive constant parameters $V_0$  and $N_0 \geq 1$. It leads to the power spectrum
\be \lb{psp}
 P_{\z}(N) = \fracmm{V^2}{12\p^2M^4_{\rm Pl}}\left( \fracmm{dV}{dN}\right)^{-1}\equiv P_0 N^2~,
\ee
and the exact scalar tilt 
\be \lb{tilt}
n_s -1 = -\fracmm{2}{N}
\ee
without any corrections with the higher powers of $N^{-1}$ on the right-hand-side in contrast to the scalar spectral index $n_s$ in the Starobinsky inflation model based on the $(R+R^2)$ modified gravity where Eq.~(\ref{tilt}) holds only approximately. 

Combining Eqs.~(\ref{r}) and (\ref{try}) yields the tensor-to-scalar ratio
\be \lb{r3}
r = \fracmm{8N_0}{N(N+N_0)}~~.
\ee 

A calculation of the inflaton scalar potential $V(\f)$ corresponding to Eq.~(\ref{try}) in the SR regime amounts to solving the differential equation
\be \lb{diff}
 \fracmm{dN}{d(\k\f)} =\sqrt{ \fracmm{N(N+N_0)}{N_0} }~~, 
\ee
where $\kappa=1/M_{\rm Pl}$. The result of integration is given by
\be \lb{rdiff}
V=V_0~\fracmm{ \cosh \left( \fracmm{\k\f}{ \sqrt{N_0}} \right) - 1} {\cosh \left( \fracmm{\k\f}{ \sqrt{N_0}} \right) + 1} 
= V_0~\tanh^2\left( \fracmm{\k\f}{2\sqrt{N_0}}\right)~~,
\ee
where the integration constant was chosen to get $V=0$ at $\f=0$.~\footnote{After a substitution $\f\to (\f-\f_0)$ it amounts to choosing $\k\f_0+\sqrt{N_0}\ln N_0=0$.}   The potential (\ref{rdiff}) is bounded from below and non-negative, $V\geq 0$. It  has the T-form with two plateaus and the Minkowski vacuum at $y=\f=0$, and it is known as the T-model of inflation in the literature,
see e.g., Refs.~\cite{Galante:2014ifa,Dalianis:2018frf,Frolovsky:2023xid}.  It is remarkable that the inflaton potential of the T-model and its scalar power spectrum take the very simple forms (\ref{try}) and (\ref{psp}) when using the e-folds variable $N$. 

Unlike the Starobinsky model, the well-known classical equivalence between the modified $F(R)$ gravity models
and the scalar-tensor gravity models in application to the T-model of inflation does not allow one to get the corresponding $F(R)$-gravity function as an elementary function. However, it is possible to get it in the leading order with respect to the first SR parameter $\ve(R)$ via replacing the pure $R^2$ term by the modulated expression,
\be \lb{moda2}
\fracmm{R^2}{8V_0}\left[ 1- 4a^2\left( \fracmm{R}{4V_0}\right)^{-2a}\right]
\approx \fracmm{R^2}{8V_0}\left[ 1- \fracmm{3}{4}\ve\right]~,
\ee
 where we have introduced $a = \sqrt{\fracmm{3}{2N_0}}$.
 
 An obvious extension of the scalar power spectrum (\ref{psp}) is given by the power-law ansatz
 \be \lb{plaw}
  P_{\z}(N) = P_0 N^{\b}
 \ee
with the real parameters $P_0>0$ and $\b>0$, which implies
\be \lb{tilt2}
n_s = 1-\fracmm{\b}{N}~~.
\ee
In this case we find the critical point at $\b=1$  without SR but SR is possible for $\b\neq 1$. For example, when $\b=3/2$ 
Eq.~(\ref{hb1}) gives rise to the scalar potential 
\be \lb{half} 
V(\f) =  6\pi^2 M^4_{\rm Pl}P_0\sqrt{N_0} \left[ \fracmm{ (\f+\f_0)^2 -8N_0M^2_{\rm Pl}}{(\f+\f_0)^2}\right]~.
\ee
Though this scalar potential is unbounded from below and is not globally defined, it does not represent a problem because its validity is limited to the SR regime on two plateaus. In the next Section, we derive a scalar potential approximately satisfying the ansatz \eqref{tilt2} with Minkowski minima.

The ansatz \eqref{plaw} leads to the scalar potential
\begin{equation}
	V=\fracmm{V_0}{1+(N/N_0)^{1-\beta}}
\end{equation}
that reduces to Eq.~\eqref{try} when $\beta=2$. The corresponding tensor-to-scalar ratio is given by
\begin{equation}\label{r_beta}
	r=\fracmm{8(\beta-1)}{N-N_0^{1-\beta}N^\beta}~.
\end{equation}

While $\beta=2$ is compatible with {\it Planck} and {\it SPT} data, the {\it ACT} data suggests a higher value \eqref{ns_P_ACT} for $n_s$, or a lower value of $\beta$, such as $\beta=3/2$ for example. Between $50$ and $60$ inflationary e-folds, $n_s$ from \eqref{tilt2} has the values
\begin{align}
	\beta &=2:~~0.96\leq n_s\leq 0.9667~,\\
	\beta &=\frac{3}{2}:~~0.97\leq n_s\leq 0.975~,
\end{align}
with the running index
\begin{equation}
\alpha_s=\fracmm{dn_s}{d\ln k}=-\fracmm{dn_s}{dN}=-\fracmm{\beta}{N^2}
\end{equation}
taking the values 
\begin{align}
	\beta &=2:~~-0.0008\leq \alpha_s\leq -0.00056~,\\
	\beta &=\frac{3}{2}:~~-0.0006\leq \alpha_s\leq -0.00042~.
\end{align}
The running index is always negative with the ansatz \eqref{tilt2}. Given the relevant values of $\beta$ as $1<\beta<2$, the
$\alpha_s$ satisfies the constraint \eqref{alphas}. The predicted values of the tensor-to-scalar ratio from Eq. \eqref{r_beta} depend upon the parameter $N_0$, while the {\it Planck} constraint $r<0.036$ implies the upper bound on $N_0$ as, for example,
\begin{align}
	\beta &=2:~~N_0<14.52~,\label{N_0_beta=2}\\
	\beta &=\frac{3}{2}:~~N_0<33.47~.\label{N_0_beta=3/2}
\end{align}

\section{Inflation in the minimal supergravity}

When inflaton is identified with the leading (single real scalar) field component $C$ of a massive vector supermultiplet, its general coupling to supergravity is governed by a real potential $J(C)$ \cite{Farakos:2013cqa,Ferrara:2013wka,Aldabergenov:2017hvp}. It is called the minimal supergravity setup for single-field inflation because there is only one physical scalar. It is equivalent to the $U(1)$ gauge-invariant model of Higgs inflation in supergravity, whose inflaton belongs to a chiral superfield charged under the $U(1)$ gauge symmetry and whose Goldstone mode is ``eaten up" by the massive gauge boson via the super-Higgs effect \cite{Aldabergenov:2017bjt,Ketov:2019toi}. For  describing inflation we only need the scalar sector of the minimal supergravity, whose Lagrangian is given by 
\begin{equation} \lb{one}
    e^{-1}\cl=\tfrac{1}{2}R-\tfrac{1}{2}J_{,CC}\partial C\partial C-\tfrac{1}{2}g^2J_{,C}^2~,
\end{equation}
with the D-type scalar potential of the non-canonical scalar field $C$ having the form $V=\tfrac{1}{2}g^2J^2_{,C}$, the gauge coupling constant $g$, and a real function $J=J(C)$. The subscripts after the commas denote the derivatives. The coupling constant $g$ determines the inflationary scale and the CMB amplitude. We take $M_{\rm Pl}=1$ for simplicity in our equations.

An impact of local supersymmetry is given by the relation between the scalar kinetic term and the
scalar potential in Eq.~(\ref{one}). As is also clear from Eq.~(\ref{one}), any scalar potential given by a real function {\it squared} can be realized in the minimal supergravity. The no-ghost condition requires $J_{,CC}>0$.

The canonical (inflalon) scalar $\phi$ can be found by integrating the equation
\begin{equation}\label{dphi_dC}
    \fracmm{d\phi}{dC}=\sqrt{J_{,CC}}~.
\end{equation}
Then the canonical inflaton potential $V(\phi)$ is obtained by inverting the solution $\phi(C)$ to Eq.~\eqref{dphi_dC} and substituting it to the potential $V$ in Eq.~\eqref{one}. The inverse function $C(\phi)$  is generically not available in the analytic form. In the special case of the Starobinsky model, the explicit solution reads
\be \lb{starJ1}
C=-\exp\left(\sqrt{\fracmm{2}{3}} \phi\right)\quad  {\rm and} \quad J= -\fracmm{3}{2}\left(\ln (-C) +C\right)~,
\ee
so that 
\be \lb{starJ2}
J_{,C}= -\fracmm{3}{2}\left(1+\fracmm{1}{C}\right) \quad {\rm and} \quad J_{,CC} 
= \fracmm{3}{2C^2}~,
\ee
leading to the canonical scalar potential $V(\phi)=\frac{9}{8}g^2\left(1-\exp\left(-\sqrt{\fracmm{2}{3}} \phi\right)\right)^2$.

It is more practical to consider the SR parameters and the inflationary observables (tilts) in terms of the non-canonical scalar $C$.  The standard potential-based SR parameters are given by
\begin{align}
    \epsilon_V &\equiv \fracmm{V_{,\phi}^2}{2V^2}=\fracmm{V^2_{,C}}{2V^2J_{,CC}}=2\fracmm{J_{,CC}}{J_{,C}^2}~,\label{epsilon_J}\\
    \eta_V &\equiv \fracmm{V_{,\phi\phi}}{V}=\fracmm{V_{,CC}}{VJ_{,CC}}-\fracmm{V_{,C}J_{,CCC}}{2VJ_{,CC}^2}=2\fracmm{J_{,CC}}{J_{,C}^2}+\fracmm{J_{,CCC}}{J_{,C}J_{,CC}}~,\label{eta_J}
\end{align}
where we have used Eqs.~(\ref{one}) and (\ref{dphi_dC}).

The total number of e-folds during inflation is given by
\begin{equation}\label{DN_J}
    N=\left|\int_{\phi_f}^{\phi_i} d\phi\fracmm{V}{V_{,\phi}}\right|=\left|\int_{C_f}^{C_i} dCJ_{,C}\right|=\fracmm{1}{2}|J(C_f)-J(C_i)|~.
\end{equation}
Accordingly,  the power spectrum of scalar perturbations, its spectral tilt $n_s$, and the tensor-to-scalar ratio $r$ during SR are
\begin{align}
    P_\zeta &\simeq \fracmm{V^3}{12\pi^2V^2_{,\phi}}=\fracmm{g^2J_{,C}^4}{96\pi^2J_{,CC}}~,\label{P_spec_J}\\
    n_s &\simeq 1+2\eta_V-6\epsilon_V=1-8\fracmm{J_{,CC}}{J_{,C}^2}+\fracmm{2J_{,CCC}}{J_{,C}J_{,CC}}~,\label{n_s_J}\\
    r &\simeq 16\epsilon_V=32\fracmm{J_{,CC}}{J_{,C}^2}~.\label{r_J}
\end{align}
In particular, the observed CMB power spectrum amplitude $P_\zeta\approx 2.1\times 10^{-9}$ \cite{Planck:2018jri} fixes the gauge coupling $g$ via Eq.~\eqref{P_spec_J}.  Equations \eqref{n_s_J} and \eqref{r_J} imply
\begin{equation}\label{n_s_rJ}
    n_s\simeq 1-\fracmm{r}{4}+\fracmm{2J_{,CCC}}{J_{,C}J_{,CC}}~.
\end{equation}

Let us assume that $J_{,CCC}=0$. Then we get $n_s\simeq 1-r/4$, and, given the {\it Planck} constraint $r<0.036$, we find $n_s>0.9804$ that is ruled out by CMB observations because, for example, the {\it P+ACT} constraint on $r<0.038$ gives rise to $n_s>0.9905$. Therefore, we have to take $J_{,CCC}\neq 0$ and  $J_{,C}J_{,CCC}<0$ at the horizon exit after taking into account that $J_{,CC}$ is necessarily positive for the correct sign of the kinetic term of $C$.

We can also express the running $\alpha_s$ of the scalar tilt purely in terms of the derivatives of $J(C)$ by using the SR approximation. For this purpose we introduce the additional SR parameter in terms of the canonical potential as
\begin{equation}\label{xi_V_def}
\xi_V\equiv\fracmm{V_{,\phi}V_{,\phi\phi\phi}}{V^2}~.
\end{equation}
Then the running index takes the form
\begin{equation}
	\alpha_s\simeq 16\epsilon_V\eta_V-24\epsilon_V^2- 2\xi_V~.
\end{equation}
After reformulating the SR parameters in terms of the derivatives of $J(C)$, we find
\begin{equation}
\alpha_s\simeq 20\fracmm{J_{,CCC}}{J_{,C}^3}-32\fracmm{J_{,CC}^2}{J_{,C}^4}+\fracmm{4J_{,CCC}^2}{J_{,C}^2J_{,CC}^2}-\fracmm{4J_{,CCCC}}{J_{,C}^2J_{,CC}}~.
\end{equation}
This includes the fourth derivative of $J(C)$, which can be related to the curvature of the K\"ahler manifold corresponding to the equivalent formulation of the $J(C)$ supergravity in terms if a chiral and a massless vector superfield \cite{Aldabergenov:2017bjt,Ketov:2019toi}.

\subsection{Reconstruction of $J$-function}

In this Subsection we reconstruct the $J$-function from the general ansatz \eqref{tilt2}, which should allow us to find $J(N)$. By using the same ansatz, the solution $C(N)$ needs to be found and inverted in order to find $J(N(C))$. We find that an exact function $J(C)$ cannot be analytically obtained from this procedure even for $\beta=2$, though some analytic approximations are possible in certain regimes.

When $N$ is the number of e-folds counted backwards, we have $\dot N=-H$, where $H$ is Hubble function. Then the equation of motion for $C(N)$ takes the form
\begin{equation}
    C_{,NN}-(3+\epsilon_H)C_{,N}+\fracmm{V_{,C}}{H^2J_{,CC}}=0~.
\end{equation}
During SR, this can be simplified to
\begin{equation}\label{C_EOM_SR}
    C_{,N}\simeq\fracmm{V_{,C}}{VJ_{,CC}}=\fracmm{2}{J_{,C}}~,
\end{equation}
where we have used the Friedmann equation $3H^2\simeq V=\tfrac{1}{2}g^2J_{,C}^2$. In terms of the real function $J(N)$, Eq.~\eqref{C_EOM_SR} is greatly simplified to
\begin{equation}\label{J(N)_sol}
    J_{,N}\simeq 2 \quad {\rm and,~hence,} \quad J(N)\simeq J_0+2(N-N_0)~,
\end{equation}
where the integration constant was chosen to get $J=J_0$ when $N=N_0$. Here $N_0$ is the same integration constant as in the preceding Section, while it should not be  confused with the initial or final e-folds values for inflation. 

The universal SR solution \eqref{J(N)_sol} is consistent with Eq.~\eqref{DN_J} and does not depend upon further details about the power spectrum, which are hidden in the relation between $N$ and $C$.  For instance, let us take the {\it ansatz} for $n_s$ from the preceding Section,
\begin{equation}\label{n_s_ansatz}
    n_s=1-\fracmm{\beta}{N}~~, \quad 1 <\beta <2~.
\end{equation}
By using $J(N)$ from \eqref{J(N)_sol}, the SR parameters \eqref{epsilon_J} and \eqref{eta_J} are given by
\begin{equation}
    \epsilon_V\simeq -\fracmm{C_{,NN}}{C_{,N}}~,~~~\eta_V\simeq \fracmm{C_{,NNN}}{2C_{,NN}}-\fracmm{5 C_{,NN}}{2C_{,N}}~,
\end{equation}
where $C_{,N}C_{,NN}<0$ is required by positivity of $J_{,CC}\simeq -2C_{,NN}/C_{,N}^3$. With these SR parameters, we find
\begin{equation}
    n_s\simeq 1+\partial_N\ln(C_{,N}C_{,NN}) \quad {\rm and} \quad r\simeq -16\fracmm{C_{,NN}}{C_{,N}}~.
\end{equation}

Next, by using Eqs.~\eqref{n_s_ansatz}, the general solution for $C_{,N}$ takes the form $C_{,N}=\pm\sqrt{b_2+b_1N^{1-\beta}}$, where the integration constants $b_1$ and $b_2$ are assumed to be positive  because, otherwise, it would lead to unrealistically large values of $r$. Integrating that relation once more yields
\begin{align}
\begin{aligned}\label{C_N_sol}
    C &=C_0\pm\int dN\sqrt{b_2+b_1N^{1-\beta}}\\
    &=C_0\pm\sqrt{b_2}N\times\,_2F_1\Big(-\frac{1}{2},\frac{1}{1-\beta},\frac{2-\beta}{1-\beta},-\frac{b_1}{b_2}N^{1-\beta}\Big)~,
\end{aligned}
\end{align}
where $_2F_1$ is the hypergeometric function, $C_0$ is another integration constant, and $\beta$ lies between $1<\beta<2$ in order to match the observed values of $n_s$. Given the solution \eqref{C_N_sol}, the tensor-to-scalar ratio $r(N)$ reads
\begin{equation}\label{r_b1b2}
    r=\fracmm{8(\beta-1)}{N+b_2N^{\beta}/b_1}~,
\end{equation}
whose parameters can be chosen to match the observational bounds on $r$. More specifically, Eq.~\eqref{r_b1b2} matches 
Eq.~\eqref{r_beta} from the preceding Section if $b_1/b_2=N_0^{\beta-1}$. By using this relation, the ratio $b_1/b_2$ is constrained by the CMB data according to Eqs.~\eqref{N_0_beta=2} and \eqref{N_0_beta=3/2}.

A derivation of  $J(C)\simeq 2(N(C)-N_0)+J_0$ requires knowing the inverse function $N(C)$ from Eq.~\eqref{C_N_sol} but it is not possible explicitly. Nevertheless, it is possible to approximate $C(N)$ in  some specific regimes where it can be analytically inverted. To do this, it is convenient to rewrite the first line of Eq.~\eqref{C_N_sol} as
\begin{equation}\label{C_N_sol2}
	C=C_0\pm \sqrt{b_2}\int dN\sqrt{1+(N/N_0)^{1-\beta}}~,
\end{equation}
where we have used $b_1/b_2=N_0^{\beta-1}$. 

Let us consider two opposite limits, $(N/N_0)^{1-\beta}\ll 1$ and $(N/N_0)^{1-\beta}\gg 1$. The former case in the leading order gives rise to
\begin{equation}
	C-C_0\simeq\pm\sqrt{b_2}(N-N_0)
\end{equation}
and, therefore, from \eqref{J(N)_sol} we get
\begin{equation}\label{J_sol_1}
	J(C)\simeq J_0\mp \fracmm{2}{\sqrt{b_2}}(C-C_0)~.
\end{equation}
In the latter case, the $C(N)$ can be approximated as
\begin{equation}
	C-C_0\simeq \pm\fracmm{2\sqrt{b_2}}{3-\beta}N_0^{\frac{\beta-1}{2}}\Big(N^{\frac{3-\beta}{2}}-N_0^{\frac{3-\beta}{2}}\Big)~,
\end{equation}
which leads to
\begin{equation}
	J(C)\simeq J_0-2N_0+2\Big[N_0^{\frac{3-\beta}{2}}\mp\fracmm{3-\beta}{2\sqrt{b_2}}N_0^{\frac{1-\beta}{2}}(C-C_0)\Big]^{\frac{2}{3-\beta}}~.
\end{equation}

The upper bounds on $N_0$ in Eqs.~\eqref{N_0_beta=2} and \eqref{N_0_beta=3/2}, derived from the observational bound on $r$, allow us to choose smaller $N_0$. Then the condition $(N/N_0)^{1-\beta}\ll 1$ holds during the most of inflation, while the linear function $J(C)$ in Eq.~\eqref{J_sol_1} becomes a good approximation. 

The leading correction to Eq.~\eqref{J_sol_1} can be computed for $\beta=3/2$ by including the next term in the expansion of the integrand in Eq.~\eqref{C_N_sol2}. Inverting the solution $C(N)$ yields~\footnote{When $\beta=2$, the function $C(N)$ beyond the leading order is not analytically invertible. 
Choosing $\beta=3/2$ can be justified by the ACT observational evidence suggesting larger values on $n_s$ when compared to 
the {\it Planck} data.}
\begin{equation}
	N-N_0\simeq \fracmm{3N_0}{2}+\fracmm{s}{\sqrt{b_2}}(C-C_0)-N_0\bigg[\fracmm{9}{4}+\fracmm{s}{\sqrt{b_2}N_0}(C-C_0)\bigg]^{1/2}~,
\end{equation}
where $s=\pm 1$ is the sign of the second term in Eq.~\eqref{C_N_sol2}. This leads to
\begin{equation}\label{J(C)_approx}
	J\simeq J_0+3N_0+\fracmm{2s}{\sqrt{b_2}}(C-C_0)-2N_0\bigg[\fracmm{9}{4}+\fracmm{s}{\sqrt{b_2}N_0}(C-C_0)\bigg]^{1/2}~.
\end{equation}

\subsection{Reconstruction-motivated model of inflation}

In this Subsection, it is demonstrated that Eq.~\eqref{J(C)_approx} can be the {\it exact} ~function $J(C)$ in a new inflation model with two Minkowski minima at finite values of $C$, without violating the no-ghost condition.

The $J$-function in Eq.~\eqref{J(C)_approx} can be rewritten to
\begin{equation}\label{J_gamma}
	J=\gamma_1 C+\gamma_2\sqrt{\gamma_3+C}~,
\end{equation}
where the new parameters $\gamma_{1,2,3}$ are related to the parameters in Eq.~\eqref{J(C)_approx} as~\footnote{We have choosen $s=+1$ and have  ignored the constant part of $J$ because they are irrelevant here.}
\begin{equation}\label{gammas}
	\gamma_1=\fracmm{2}{\sqrt{b_2}}~,~~~\gamma_2=-2\fracmm{\sqrt{N_0}}{b_2^{1/4}}~,~~~\gamma_3=\fracmm{9}{4}\sqrt{b_2}N_0-C_0~.
\end{equation}

The non-canonical scalar $C$ in the model \eqref{J_gamma} is related to the canonical scalar $\phi$ as follows:
\begin{equation}\label{C_canonical}
	C=\fracmm{\phi^4}{16\gamma_2^2}-\gamma_3=\fracmm{\sqrt{b_2}\phi^4}{64N_0}-\fracmm{9}{4}\sqrt{b_2}N_0+C_0~,
\end{equation}
which yields the canonical scalar potential
\begin{equation}\label{V_gamma}
	V=\fracmm{1}{2}g^2J_{,C}^2=2\fracmm{g^2}{b_2}\bigg(1-\fracmm{4N_0}{\phi^2}\bigg)^2~,
\end{equation}
where we have used Eqs.~\eqref{J_gamma}, \eqref{gammas} and \eqref{C_canonical}. Therefore, the parameters $b_2$, $N_0$ and the gauge coupling $g$ have direct physical meaning. In particular, the ratio $g^2/b_2$ fixes the amplitude of scalar perturbations, and the $N_0$  determines the observables $n_s$, $r$, and $\alpha_s$. The potential (\ref{V_gamma}) is 
well-behaved because the singularity at $\phi=0$ is screened by the infinite walls, while the inflaton field $\phi$ has the non-vanishing vacuum expectation values in two Minkowski minima.

\begin{figure*}
\centering
  \centering
  \includegraphics[width=.65\linewidth]{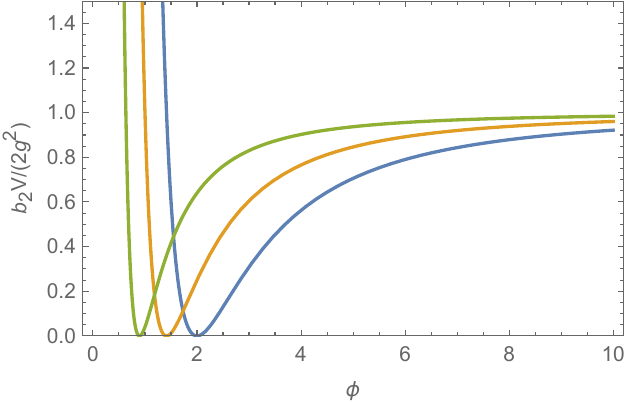}
\captionsetup{width=1\linewidth}
\caption{The potential $b_2V/(2g^2)$ from Eq. \eqref{V_gamma} with $N_0=1$ (blue curve), $N_0=1/2$ (orange) and $N_0=1/5$ (green).}\label{Fig_potential}
\end{figure*}

The potential \eqref{V_gamma} is shown in Fig.~\ref{Fig_potential}. It is symmetric with respect to the sign flip of $\phi$, has the singularity at $\phi=0$, and two stable Minkowski vacua at $\langle\phi\rangle=\pm 2\sqrt{N_0}$. Slow-roll inflation takes place for large $|\phi|$, where the potential approaches the constant value $2g^2/b_2$. The potential \eqref{V_gamma} is essentially a square of the potential \eqref{half} with $\beta=3/2$.

The number of e-folds from the horizon exit in the model \eqref{V_gamma} is given by
\begin{equation}\label{N_gamma}
	N\simeq\bigg(\fracmm{\phi^4}{64N_0}-\fracmm{\phi^2}{8}\bigg)\bigg|^{\phi_*}_{\phi_e}~,
\end{equation}
where $\phi_e$ is the value at the end of inflation, which is usually taken at $\epsilon_V=1$. 

The condition $\epsilon_V=1$ yields a cubic equation for $\phi_e$ but we can simplify this task by taking $\phi_e=\langle\phi\rangle$, which  is justified because the number of e-folds between the event at $\epsilon_V=1$ and 
the one at $\phi=\langle\phi\rangle$ is small when compared to the duration of inflation. Given $\phi_e=\langle\phi\rangle=2\sqrt{N_0}$, Eq.~\eqref{N_gamma} yields
\begin{equation}
	N\simeq\fracmm{\phi_*^4}{64N_0}-\fracmm{\phi_*^2}{8}+\fracmm{N_0}{4}~.
\end{equation}
After inverting this equation, we obtain $\phi_*(N)$ as
\begin{equation}
	\phi_*\simeq 2\sqrt{N_0(1+2\sqrt{N/N_0})}~.
\end{equation}
By using this $\phi_*(N)$, we can estimate the inflationary observables as
\begin{gather}\label{Obs_gamma}
\begin{gathered}
	n_s\simeq 1-\fracmm{2+3\sqrt{N/N_0}}{N(1+2\sqrt{N/N_0})}~,~~~r\simeq\fracmm{8}{N(1+2\sqrt{N/N_0})}~,\\
	\alpha_s\simeq\fracmm{1-3(1+2\sqrt{N/N_0})(3+4\sqrt{N/N_0})}{4N^2(1+2\sqrt{N/N_0})^2}~.
\end{gathered}
\end{gather}

The spectral tilt ansatz \eqref{tilt2} with $\beta=3/2$ is reproduced for large values of $\sqrt{N/N_0}$. In this
case, Eq.~(\ref{Obs_gamma}) gets simplified to
\begin{equation}\label{obs_limit}
	n_s\simeq 1-\fracmm{3}{2N}~,~~~r\simeq\fracmm{4}{N\sqrt{N/N_0}}~,~~~\alpha_s\simeq -\fracmm{3}{2N^2}~.
\end{equation}

\begin{figure*}
\centering
  \centering
  \includegraphics[width=1\linewidth]{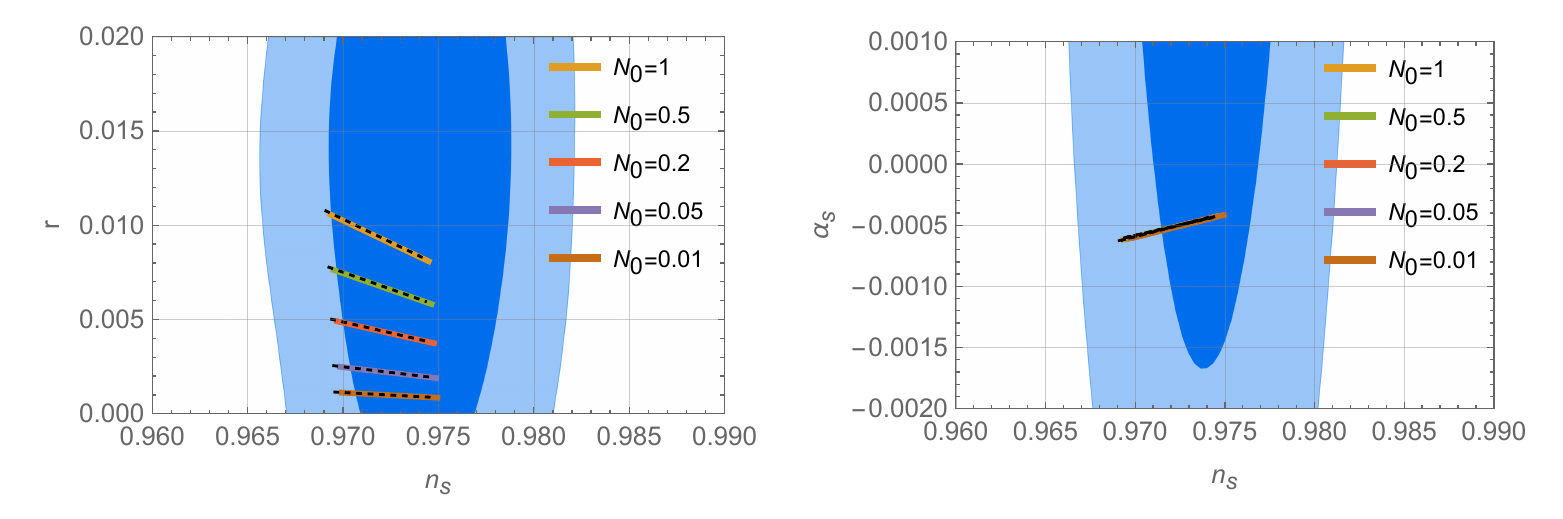}
\captionsetup{width=1\linewidth}
\caption{The inflationary observables $n_s$, $r$, and $\alpha_s$ obtained from Eq.~\eqref{Obs_gamma} for $N=50\div 60$. The dashed black lines show the results of numerical integration. The observational constraints are taken from {\it Planck+ACT} data \cite{ACT:2025tim}.}\label{Fig_obs}
\end{figure*}

The plots of $n_s$, $r$, and $\alpha_s$ from Eq.~\eqref{Obs_gamma} are shown in Fig.~\ref{Fig_obs} for a few different values of $N_0$ with $N$ between $50$ and $60$ e-folds, where larger e-folds lead to larger $n_s$. The observational constraints are taken from the combined {\it Planck+ACT} data \cite{ACT:2025tim} with the help of {\it GetDist} package \cite{Lewis:2019xzd}. For comparison, the results from numerical integration of the equations of motion are shown by the dashed black lines. As is expected from Eq.~\eqref{obs_limit} and is confirmed by Fig.~\ref{Fig_obs}, the tensor-to-scalar ratio $r$ is significantly dependent upon $N_0$, while the scalar tilt $n_s$ and its running $\alpha_s$ are largely independent of $N_0$, thus providing the robust predictions.

It is worth noticing that the potential \eqref{V_gamma} can describe both large-field or small-field inflation depending upon the value of $N_0$. For example, for small enough $N_0$, the distance traveled by inflaton during inflation can be subPlanckian
when
\begin{equation}
	\phi_*-\langle\phi\rangle=2\sqrt{2\sqrt{N_0N}} -2\sqrt{N_0}<1~.
\end{equation}

\section{Conclusion}
\label{Sec_concl}

The minimal supergravity framework provides a fundamental motivation for the inflation model building, while it is highly restrictive. Nevertheless,  as is demonstrated above, the minimal supergravity models can accommodate most recent (precision) CMB observations. As an illustration, we provided two new viable inflation models in this framework, the one motivated by simplicity and another one motivated by reconstruction. In our models, reheating can be added via supergravity coupling to supersymmetric matter along the standard lines, whereas spontaneous supersymmetry breaking after inflation cannot be achieved and requires adding a hidden sector, see Refs.~\cite{Aldabergenov:2017hvp,Aldabergenov:2018nzd} for details.

\section*{Acknowledgements}
S.V.K. was partially supported by Tokyo Metropolitan University, Tomsk State University under the development program Priority-2030, and  the World Premier International Research Center Initiative, MEXT, Japan. The authors are grateful to Chengjie Fu, Constantinos Pallis
and Swapnil Kumar Singh for correspondence.

\providecommand{\href}[2]{#2}\begingroup\raggedright\endgroup


\begin{thebibliography}{10}

\bibitem{Martin:2024qnn}
J.~Martin, C.~Ringeval, and V.~Vennin, ``{Cosmic Inflation at the
  crossroads},'' \href{http://dx.doi.org/10.1088/1475-7516/2024/07/087}{{\em
  JCAP} {\bfseries 07} (2024) 087},
  \href{http://arxiv.org/abs/2404.10647}{{\ttfamily arXiv:2404.10647
  [astro-ph.CO]}}.

\bibitem{Gong:2025pbi}
J.-O. Gong, S.~V. Ketov, and T.~Terada, ``{$F$-term Multi-Field Inflation in
  Supergravity without Stabiliser Superfields},''
  \href{http://arxiv.org/abs/2508.20194}{{\ttfamily arXiv:2508.20194
  [hep-th]}}.

\bibitem{Farakos:2013cqa}
F.~Farakos, A.~Kehagias, and A.~Riotto, ``{On the Starobinsky Model of
  Inflation from Supergravity},''
  \href{http://dx.doi.org/10.1016/j.nuclphysb.2013.08.005}{{\em Nucl. Phys. B}
  {\bfseries 876} (2013) 187--200},
  \href{http://arxiv.org/abs/1307.1137}{{\ttfamily arXiv:1307.1137 [hep-th]}}.

\bibitem{Ferrara:2013wka}
S.~Ferrara, R.~Kallosh, and A.~Van~Proeyen, ``{On the Supersymmetric Completion
  of $R+R^2$ Gravity and Cosmology},''
  \href{http://dx.doi.org/10.1007/JHEP11(2013)134}{{\em JHEP} {\bfseries 11}
  (2013) 134}, \href{http://arxiv.org/abs/1309.4052}{{\ttfamily arXiv:1309.4052
  [hep-th]}}.

\bibitem{Aldabergenov:2017hvp}
Y.~Aldabergenov and S.~V. Ketov, ``{Removing instability of inflation in
  Polonyi{\textendash}Starobinsky supergravity by adding FI term},''
  \href{http://dx.doi.org/10.1142/S0217732318500323}{{\em Mod. Phys. Lett. A}
  {\bfseries 33} no.~5, (2018) 1850032},
  \href{http://arxiv.org/abs/1711.06789}{{\ttfamily arXiv:1711.06789
  [hep-th]}}.

\bibitem{Planck:2018jri}
{\bfseries Planck} Collaboration, Y.~Akrami {\em et~al.}, ``{Planck 2018
  results. X. Constraints on inflation},''
  \href{http://dx.doi.org/10.1051/0004-6361/201833887}{{\em Astron. Astrophys.}
  {\bfseries 641} (2020) A10},
  \href{http://arxiv.org/abs/1807.06211}{{\ttfamily arXiv:1807.06211
  [astro-ph.CO]}}.

\bibitem{BICEP:2021xfz}
{\bfseries BICEP, Keck} Collaboration, P.~A.~R. Ade {\em et~al.}, ``{Improved
  Constraints on Primordial Gravitational Waves using Planck, WMAP, and
  BICEP/Keck Observations through the 2018 Observing Season},''
  \href{http://dx.doi.org/10.1103/PhysRevLett.127.151301}{{\em Phys. Rev.
  Lett.} {\bfseries 127} no.~15, (2021) 151301},
  \href{http://arxiv.org/abs/2110.00483}{{\ttfamily arXiv:2110.00483
  [astro-ph.CO]}}.

\bibitem{ACT:2025tim}
{\bfseries ACT} Collaboration, E.~Calabrese {\em et~al.}, ``{The Atacama
  Cosmology Telescope: DR6 Constraints on Extended Cosmological Models},''
  \href{http://arxiv.org/abs/2503.14454}{{\ttfamily arXiv:2503.14454
  [astro-ph.CO]}}.

\bibitem{SPT-3G:2025bzu}
{\bfseries SPT-3G} Collaboration, E.~Camphuis {\em et~al.}, ``{SPT-3G D1: CMB
  temperature and polarization power spectra and cosmology from 2019 and 2020
  observations of the SPT-3G Main field},''
  \href{http://arxiv.org/abs/2506.20707}{{\ttfamily arXiv:2506.20707
  [astro-ph.CO]}}.

\bibitem{Dioguardi:2025vci}
C.~Dioguardi, A.~J. Iovino, and A.~Racioppi, ``{Fractional attractors in light
  of the latest ACT observations},''
  \href{http://dx.doi.org/10.1016/j.physletb.2025.139664}{{\em Phys. Lett. B}
  {\bfseries 868} (2025) 139664},
  \href{http://arxiv.org/abs/2504.02809}{{\ttfamily arXiv:2504.02809 [gr-qc]}}.

\bibitem{Kallosh:2025rni}
R.~Kallosh, A.~Linde, and D.~Roest, ``{ACT, SPT, and chaotic inflation},''
  \href{http://arxiv.org/abs/2503.21030}{{\ttfamily arXiv:2503.21030
  [hep-th]}}.

\bibitem{Gialamas:2025ofz}
I.~D. Gialamas, T.~Katsoulas, and K.~Tamvakis, ``{Keeping the relation between
  the Starobinsky model and no-scale supergravity ACTive},''
  \href{http://dx.doi.org/10.1088/1475-7516/2025/09/060}{{\em JCAP} {\bfseries
  09} (2025) 060}, \href{http://arxiv.org/abs/2505.03608}{{\ttfamily
  arXiv:2505.03608 [gr-qc]}}.

\bibitem{Aoki:2025wld}
S.~Aoki, H.~Otsuka, and R.~Yanagita, ``{Higgs-modular inflation},''
  \href{http://dx.doi.org/10.1103/v4z9-676d}{{\em Phys. Rev. D} {\bfseries 112}
  no.~4, (2025) 043505}, \href{http://arxiv.org/abs/2504.01622}{{\ttfamily
  arXiv:2504.01622 [hep-ph]}}.

\bibitem{Berera:2025vsu}
A.~Berera, S.~Brahma, Z.~Qiu, R.~O.~Ramos, and G.~S. Rodrigues, ``{The early
  universe is $\textit{ACT}$-ing $\textit{warm}$},''
  \href{http://arxiv.org/abs/2504.02655}{{\ttfamily arXiv:2504.02655
  [hep-th]}}.

\bibitem{Brahma:2025dio}
S.~Brahma and J.~Calder\'on-Figueroa, ``{Is the CMB revealing signs of
  pre-inflationary physics?},''
  \href{http://arxiv.org/abs/2504.02746}{{\ttfamily arXiv:2504.02746
  [astro-ph.CO]}}.

\bibitem{Gialamas:2025kef}
I.~D. Gialamas, A.~Karam, A.~Racioppi, and M.~Raidal, ``{Has ACT measured
  radiative corrections to the tree-level Higgs-like inflation?},''
  \href{http://arxiv.org/abs/2504.06002}{{\ttfamily arXiv:2504.06002
  [astro-ph.CO]}}.

\bibitem{Salvio:2025izr}
A.~Salvio, ``{Independent connection in action during inflation},''
  \href{http://dx.doi.org/10.1103/tq3v-vy3y}{{\em Phys. Rev. D} {\bfseries 112}
  no.~6, (2025) L061301}, \href{http://arxiv.org/abs/2504.10488}{{\ttfamily
  arXiv:2504.10488 [hep-ph]}}.

\bibitem{Antoniadis:2025pfa}
I.~Antoniadis, J.~Ellis, W.~Ke, D.~V. Nanopoulos, and K.~A. Olive, ``{How
  accidental was inflation?},''
  \href{http://dx.doi.org/10.1088/1475-7516/2025/08/090}{{\em JCAP} {\bfseries
  08} (2025) 090}, \href{http://arxiv.org/abs/2504.12283}{{\ttfamily
  arXiv:2504.12283 [hep-ph]}}.

\bibitem{Dioguardi:2025mpp}
C.~Dioguardi and A.~Karam, ``{Palatini Linear Attractors Are Back in ACTion},''
  \href{http://arxiv.org/abs/2504.12937}{{\ttfamily arXiv:2504.12937 [gr-qc]}}.

\bibitem{Kuralkar:2025zxr}
H.~J. Kuralkar, S.~Panda, and A.~Vidyarthi, ``{Effective Starobinsky
  pre-inflation},'' \href{http://arxiv.org/abs/2504.15061}{{\ttfamily
  arXiv:2504.15061 [gr-qc]}}.

\bibitem{Gao:2025onc}
Q.~Gao, Y.~Gong, Z.~Yi, and F.~Zhang, ``{Nonminimal coupling in light of ACT
  data},'' \href{http://dx.doi.org/10.1016/j.dark.2025.102106}{{\em Phys. Dark
  Univ.} {\bfseries 50} (2025) 102106},
  \href{http://arxiv.org/abs/2504.15218}{{\ttfamily arXiv:2504.15218
  [astro-ph.CO]}}.

\bibitem{He:2025bli}
M.~He, M.~Hong, and K.~Mukaida, ``{Increase of ns in regularized pole inflation
  {\&} Einstein-Cartan gravity},''
  \href{http://dx.doi.org/10.1088/1475-7516/2025/09/080}{{\em JCAP} {\bfseries
  09} (2025) 080}, \href{http://arxiv.org/abs/2504.16069}{{\ttfamily
  arXiv:2504.16069 [astro-ph.CO]}}.

\bibitem{Drees:2025ngb}
M.~Drees and Y.~Xu, ``{Refined predictions for Starobinsky inflation and
  post-inflationary constraints in light of ACT},''
  \href{http://dx.doi.org/10.1016/j.physletb.2025.139612}{{\em Phys. Lett. B}
  {\bfseries 867} (2025) 139612},
  \href{http://arxiv.org/abs/2504.20757}{{\ttfamily arXiv:2504.20757
  [astro-ph.CO]}}.

\bibitem{Haque:2025uri}
M.~R. Haque, S.~Pal, and D.~Paul, ``{ACT DR6 Insights on the Inflationary
  Attractor models and Reheating},''
  \href{http://arxiv.org/abs/2505.01517}{{\ttfamily arXiv:2505.01517
  [astro-ph.CO]}}.

\bibitem{Liu:2025qca}
L.~Liu, Z.~Yi, and Y.~Gong, ``{Reconciling Higgs Inflation with ACT
  Observations through Reheating},''
  \href{http://arxiv.org/abs/2505.02407}{{\ttfamily arXiv:2505.02407
  [astro-ph.CO]}}.

\bibitem{Addazi:2025qra}
A.~Addazi, Y.~Aldabergenov, and S.~V. Ketov, ``{Curvature corrections to
  Starobinsky inflation can explain the ACT results},''
  \href{http://dx.doi.org/10.1016/j.physletb.2025.139883}{{\em Phys. Lett. B}
  {\bfseries 869} (2025) 139883},
  \href{http://arxiv.org/abs/2505.10305}{{\ttfamily arXiv:2505.10305 [gr-qc]}}.

\bibitem{Byrnes:2025kit}
C.~T. Byrnes, M.~Cort\^es, and A.~R. Liddle, ``{The curvaton ACTs again},''
  \href{http://arxiv.org/abs/2505.09682}{{\ttfamily arXiv:2505.09682
  [astro-ph.CO]}}.

\bibitem{Yogesh:2025wak}
Yogesh, A.~Mohammadi, Q.~Wu, and T.~Zhu, ``{Starobinsky like inflation and EGB
  Gravity in the light of ACT},''
  \href{http://arxiv.org/abs/2505.05363}{{\ttfamily arXiv:2505.05363
  [astro-ph.CO]}}.

\bibitem{Kallosh:2025ijd}
R.~Kallosh and A.~Linde, ``{On the Present Status of Inflationary Cosmology},''
  \href{http://arxiv.org/abs/2505.13646}{{\ttfamily arXiv:2505.13646
  [hep-th]}}.

\bibitem{Kohri:2025lau}
K.~Kohri and X.~Wang, ``{Primordial Black Holes Save $R^2$ Inflation},''
  \href{http://arxiv.org/abs/2506.06797}{{\ttfamily arXiv:2506.06797
  [astro-ph.CO]}}.

\bibitem{Ahmad:2025mul}
M.~N. Ahmad and M.~U. Rehman, ``{Supersymmetric Hybrid Inflation with
  K{\"a}hler-Induced $\mathbf{R}$-Symmetry Breaking},''
  \href{http://arxiv.org/abs/2506.23244}{{\ttfamily arXiv:2506.23244
  [hep-ph]}}.

\bibitem{Okada:2025lpl}
N.~Okada and O.~Seto, ``{Smooth hybrid inflation in light of ACT DR6 data},''
  \href{http://arxiv.org/abs/2506.15965}{{\ttfamily arXiv:2506.15965
  [hep-ph]}}.

\bibitem{Choudhury:2025vso}
S.~Choudhury, G.~Bauyrzhan, S.~K. Singh, and K.~Yerzhanov, ``{What new physics
  can we extract from inflation using the ACT DR6 and DESI DR2
  Observations?},'' \href{http://arxiv.org/abs/2506.15407}{{\ttfamily
  arXiv:2506.15407 [astro-ph.CO]}}.

\bibitem{Han:2025cwk}
J.~Han, H.~M. Lee, and J.-H. Song, ``{Higgs pole inflation with loop
  corrections in light of ACT results},''
  \href{http://arxiv.org/abs/2506.21189}{{\ttfamily arXiv:2506.21189
  [hep-ph]}}.

\bibitem{Leontaris:2025hly}
G.~K. Leontaris and P.~Shukla, ``{Assisted Fibre Inflation in Perturbative
  LVS},'' \href{http://arxiv.org/abs/2506.22630}{{\ttfamily arXiv:2506.22630
  [hep-th]}}.

\bibitem{McDonald:2025tfp}
J.~McDonald, ``{Unitarity-Conserving Non-Minimally Coupled Inflation and the
  ACT Spectral Index},'' \href{http://arxiv.org/abs/2506.12916}{{\ttfamily
  arXiv:2506.12916 [hep-ph]}}.

\bibitem{Gao:2025viy}
Q.~Gao, Y.~Qian, Y.~Gong, and Z.~Yi, ``{Observational constraints on
  inflationary models with non-minimally derivative coupling by ACT},''
  \href{http://dx.doi.org/10.1088/1475-7516/2025/08/083}{{\em JCAP} {\bfseries
  08} (2025) 083}, \href{http://arxiv.org/abs/2506.18456}{{\ttfamily
  arXiv:2506.18456 [gr-qc]}}.

\bibitem{Wolf:2025ecy}
W.~J. Wolf, ``{Inflationary attractors and radiative corrections in light of
  ACT},'' \href{http://arxiv.org/abs/2506.12436}{{\ttfamily arXiv:2506.12436
  [astro-ph.CO]}}.

\bibitem{Singh:2025uyr}
S.~K. Singh, ``{Symmetry-Protected $α$-Attractor Hybrid Inflation in
  Supergravity and Constraints from ACT DR6 and DESI DR2},''
  \href{http://arxiv.org/abs/2511.05545}{{\ttfamily arXiv:2511.05545
  [hep-ph]}}.

\bibitem{Fu:2025ciy}
C.~Fu, D.~Lu, and S.-J. Wang, ``{The Harrison-Zeldovich attractor: From Planck
  to ACT},'' \href{http://arxiv.org/abs/2510.24682}{{\ttfamily arXiv:2510.24682
  [astro-ph.CO]}}.

\bibitem{Ahmed:2025eip}
W.~Ahmed, C.~Pallis, and M.~U. Rehman, ``{GUT-Scale Smooth Hybrid Inflation
  with a Stabilized Modulus in Light of ACT and SPT Data},''
  \href{http://arxiv.org/abs/2510.20478}{{\ttfamily arXiv:2510.20478
  [hep-ph]}}.

\bibitem{Hodges:1990bf}
H.~M. Hodges and G.~R. Blumenthal, ``{Arbitrariness of inflationary fluctuation
  spectra},'' \href{http://dx.doi.org/10.1103/PhysRevD.42.3329}{{\em Phys. Rev.
  D} {\bfseries 42} (1990) 3329--3333}.

\bibitem{Galante:2014ifa}
M.~Galante, R.~Kallosh, A.~Linde, and D.~Roest, ``{Unity of Cosmological
  Inflation Attractors},''
  \href{http://dx.doi.org/10.1103/PhysRevLett.114.141302}{{\em Phys. Rev.
  Lett.} {\bfseries 114} no.~14, (2015) 141302},
  \href{http://arxiv.org/abs/1412.3797}{{\ttfamily arXiv:1412.3797 [hep-th]}}.

\bibitem{Dalianis:2018frf}
I.~Dalianis, A.~Kehagias, and G.~Tringas, ``{Primordial black holes from
  {\ensuremath{\alpha}}-attractors},''
  \href{http://dx.doi.org/10.1088/1475-7516/2019/01/037}{{\em JCAP} {\bfseries
  01} (2019) 037}, \href{http://arxiv.org/abs/1805.09483}{{\ttfamily
  arXiv:1805.09483 [astro-ph.CO]}}.

\bibitem{Frolovsky:2023xid}
D.~Frolovsky and S.~V. Ketov, ``{Fitting Power Spectrum of Scalar Perturbations
  for Primordial Black Hole Production during Inflation},''
  \href{http://dx.doi.org/10.3390/astronomy2010005}{{\em Astronomy} {\bfseries
  2} no.~1, (2023) 47--57}, \href{http://arxiv.org/abs/2302.06153}{{\ttfamily
  arXiv:2302.06153 [astro-ph.CO]}}.

\bibitem{Aldabergenov:2017bjt}
Y.~Aldabergenov and S.~V. Ketov, ``{Higgs mechanism and cosmological constant
  in $N=1$ supergravity with inflaton in a vector multiplet},''
  \href{http://dx.doi.org/10.1140/epjc/s10052-017-4807-8}{{\em Eur. Phys. J. C}
  {\bfseries 77} no.~4, (2017) 233},
  \href{http://arxiv.org/abs/1701.08240}{{\ttfamily arXiv:1701.08240
  [hep-th]}}.

\bibitem{Ketov:2019toi}
S.~V. Ketov, ``{On the equivalence of Starobinsky and Higgs inflationary models
  in gravity and supergravity},''
  \href{http://dx.doi.org/10.1088/1751-8121/ab6a33}{{\em J. Phys. A} {\bfseries
  53} no.~8, (2020) 084001}, \href{http://arxiv.org/abs/1911.01008}{{\ttfamily
  arXiv:1911.01008 [hep-th]}}.

\bibitem{Lewis:2019xzd}
A.~Lewis, ``{GetDist: a Python package for analysing Monte Carlo samples},''
  \href{http://dx.doi.org/10.1088/1475-7516/2025/08/025}{{\em JCAP} {\bfseries
  08} (2025) 025}, \href{http://arxiv.org/abs/1910.13970}{{\ttfamily
  arXiv:1910.13970 [astro-ph.IM]}}.

\bibitem{Aldabergenov:2018nzd}
Y.~Aldabergenov, S.~V. Ketov, and R.~Knoops, ``{General couplings of a vector
  multiplet in $N=1$ supergravity with new FI terms},''
  \href{http://dx.doi.org/10.1016/j.physletb.2018.07.072}{{\em Phys. Lett. B}
  {\bfseries 785} (2018) 284--287},
  \href{http://arxiv.org/abs/1806.04290}{{\ttfamily arXiv:1806.04290
  [hep-th]}}.

\end{thebibliography}
\end{document}